 \documentclass[journal]{IEEEtran}

\usepackage{makecell}
\usepackage{array}
\usepackage{graphicx,amssymb,amsmath}
\usepackage{multicol}
\usepackage[noadjust]{cite}
\usepackage{setspace}
\usepackage{subfigure}
\usepackage{graphicx}
\usepackage{float}
\usepackage {url}
\usepackage{stfloats}
\usepackage{amsthm,pifont}
\usepackage{flushend}
\usepackage{cases,subeqnarray}
\usepackage{bm,multirow,bigstrut}
\usepackage{amsmath, amsthm, amssymb}
\usepackage{textcomp}
\usepackage{latexsym,bm}
\usepackage{booktabs}
\usepackage{xcolor}
\usepackage{mathtools}
\usepackage{dsfont}
\usepackage{extarrows}
\usepackage{epsfig}
\usepackage{epsfig}
\usepackage{algpseudocode}

\usepackage{algorithmicx,algorithm}

\usepackage{cite}
\usepackage{bm}
\usepackage{multicol}       
\usepackage{multirow}       
\usepackage{array}          
\usepackage{colortbl}
\usepackage{makecell}
\definecolor{crimson}{RGB}{192,0,0}         
\definecolor{navy}{RGB}{47,85,151}         
\usepackage{bbding}
\usepackage{graphicx}
\usepackage{booktabs}
\usepackage{algorithm}
\usepackage{algpseudocode}
\usepackage{diagbox}

\usepackage{graphicx}
\usepackage{setspace}
\usepackage{cleveref}

\usepackage{url}
\usepackage{mathtools}
\def\BibTeX{{\rm B\kern-.05em{\sc i\kern-.025em b}\kern-.08em
 T\kern-.1667em\lower.7ex\hbox{E}\kern-.125emX}}

\usepackage{enumerate}

\usepackage{multicol}

\theoremstyle{plain}

\theoremstyle{plain}

\usepackage{amsmath}

\usepackage{threeparttable}
\usepackage{pdfsync}
\usepackage{flushend}

\IEEEoverridecommandlockouts

\begin{document}
\title{Near-Field User Localization and Channel Estimation for XL-MIMO Systems: Fundamentals, Recent Advances, and Outlooks \\
}

\author{Hao Lei, Jiayi Zhang,~\IEEEmembership{Senior Member,~IEEE}, Zhe Wang, Huahua Xiao,\\ Bo Ai,~\IEEEmembership{Fellow,~IEEE}, Emil Bj\"ornson,~\IEEEmembership{Fellow,~IEEE}

\thanks{H. Lei, J. Zhang, Z. Wang and B. Ai are with Beijing Jiaotong University; H. Xiao is with ZTE Corporation, State Key Laboratory of Mobile Network and Mobile Multimedia Technology; E. Bj\"ornson is with the KTH Royal Institute of Technology, Stockholm, Sweden. }

}

\maketitle

\begin{abstract}
Extremely large-scale multiple-input multiple-output (XL-MIMO) is believed to be a cornerstone of sixth-generation (6G) wireless networks. 
XL-MIMO uses more antennas to both achieve unprecedented spatial degrees of freedom (DoFs) and exploit new electromagnetic (EM) phenomena occurring in the radiative near-field.
The near-field effects provide the XL-MIMO array with depth perception, enabling precise localization and spatially multiplexing jointly in the angle and distance domains.
This article delineates the distinctions between near-field and far-field propagation, highlighting the unique EM characteristics introduced by having large antenna arrays. 
It thoroughly examines the challenges these new near-field characteristics pose for user localization and channel estimation and provides a comprehensive review of new algorithms developed to address them. The article concludes by identifying critical future research directions.


\end{abstract}


\IEEEpeerreviewmaketitle

\section{Introduction}
\label{Introduction}

As the fifth generation (5G) of wireless communication networks is being deployed worldwide, the development of the next sixth generation (6G) is steadily transitioning from a visionary to an implementation stage.
One of the cornerstone technologies of 5G is the massive multiple-input multiple-output (mMIMO) technology.
This technology equips each base station (BS) with around 64 antennas and uses them for spatially multiplexing of tens of user equipments (UEs), which significantly improves the spectral efficiency (SE) compared to classical small-scale MIMO technology \cite{[1]}, \cite{[6]}.
However, with the exponential growth of mobile user devices and their related traffic, the network technology must support even more traffic and new services with diverse and demanding requirements.
Consequently, the sixth-generation (6G) technology must be developed to support traffic volumes at least 100 times higher than those achievable in current networks.
Since network densification is commercially unattractive, each BS needs to support more traffic, which calls for using a larger number of antennas.
Hence, the basic idea of the next-generation MIMO technology revolves around deploying an extremely large number of antennas, referred to as extremely large-scale MIMO (XL-MIMO) \cite{[1]}, \cite{[6]}, \cite{[2]} or extremely large aperture arrays (ELAA). Both terms indicate that the area is large, compared to the wavelength, to significantly enhance the spatial resolution and spatial degrees of freedom (DoF).
There is also the concept of holographic MIMO (also called continuous-aperture MIMO) that squeezes in many antennas by shrinking the antenna spacing, which does not lead to the mentioned gains.

The UEs in current 5G systems are in the far-field of the mMIMO array, which has also been the case in previous network generations. However, the introduction of XL-MIMO technology featuring an extremely large aperture area opens the box to new electromagnetic (EM) characteristics that are only noticeable in the radiative near-field.
The Fraunhofer/Rayleigh distance is conventionally taken as the boundary between the near-field and far-field \cite{[1]}, \cite{[2]}. It is proportional to the squared aperture length and inversely proportional to the wavelength. In situations with physically larger arrays or smaller wavelengths than in 5G, the Fraunhofer distance can reach a kilometer, thereby making near-field propagation the norm in XL-MIMO systems rather than the exception. 
Importantly, we are not considering the reactive near-field effects that only appear very close to the array, but only how the radiated fields from a large array require long distances before becoming far-field-like. The radiative near-field is also called the Fresnel region.

The shift to near-field operation creates several new and distinctive EM characteristics. The first one is the \textbf{\emph{spherical wave properties}}, which must be considered when analyzing wireless channels. This contrasts far-field mMIMO systems, where the EM fields can be approximated as plane waves when reaching the UEs.
The second one is the \textbf{\emph{finite-depth beamfocusing}}, which allows the XL-MIMO array to focus signals on confined areas in both distance and angle, and to localize UEs in both distance and angle.
The locations obtained through user localization can be leveraged to enhance the accuracy or reduce the complexity of channel estimation. 
Finally, there will be \textbf{\emph{spatial non-stationary properties}}  due to near-field scattering and blockages. This implies that antennas positioned at different spatial locations can exhibit very distinctly different channel characteristics.


These EM characteristics are not theoretical curiosities but an opportunity for enabling new features and more efficient algorithms. The spherical curvature makes propagation channels richer \cite{[6]}, leading to more well-conditioned MIMO channels. The finite-depth beamfocusing enables spatial multiplexing of users with similar angular characteristics but different distances. It also enables a single array to perform accurate user localization \cite{[8]}; while far-field mMIMO enables angle estimation, XL-MIMO also enables distance estimation.
This capability is both useful for user localization (and sensing) and for aiding channel estimation. 
This is aligned with the general trend of integrated sensing and communication (ISAC).  

To capitalize on these new capabilities, we need to develop new near-field user localization and channel estimation algorithms customized to the new near-field EM characteristics, as well as associated hardware impairments. 
This article provides an overview of this topic. We describe the fundamental near-field properties and comprehensively review recent near-field user localization and channel estimation schemes, and describe their benefits through numerical examples. 
Finally, we identify several future research directions for near-field XL-MIMO systems, paving the way for further advancements in this emerging research area.

\section{Fundamental Characteristics of Near-Field XL-MIMO Systems}

In this section, we present three new fundamental EM characteristics of near-field XL-MIMO systems.
Notably, these characteristics delineate the main disparities between far-field and near-field communication, concurrently presenting novel challenges and opportunities within near-field XL-MIMO systems.

\begin{figure}
  \centering
  \includegraphics[width=3.5in]{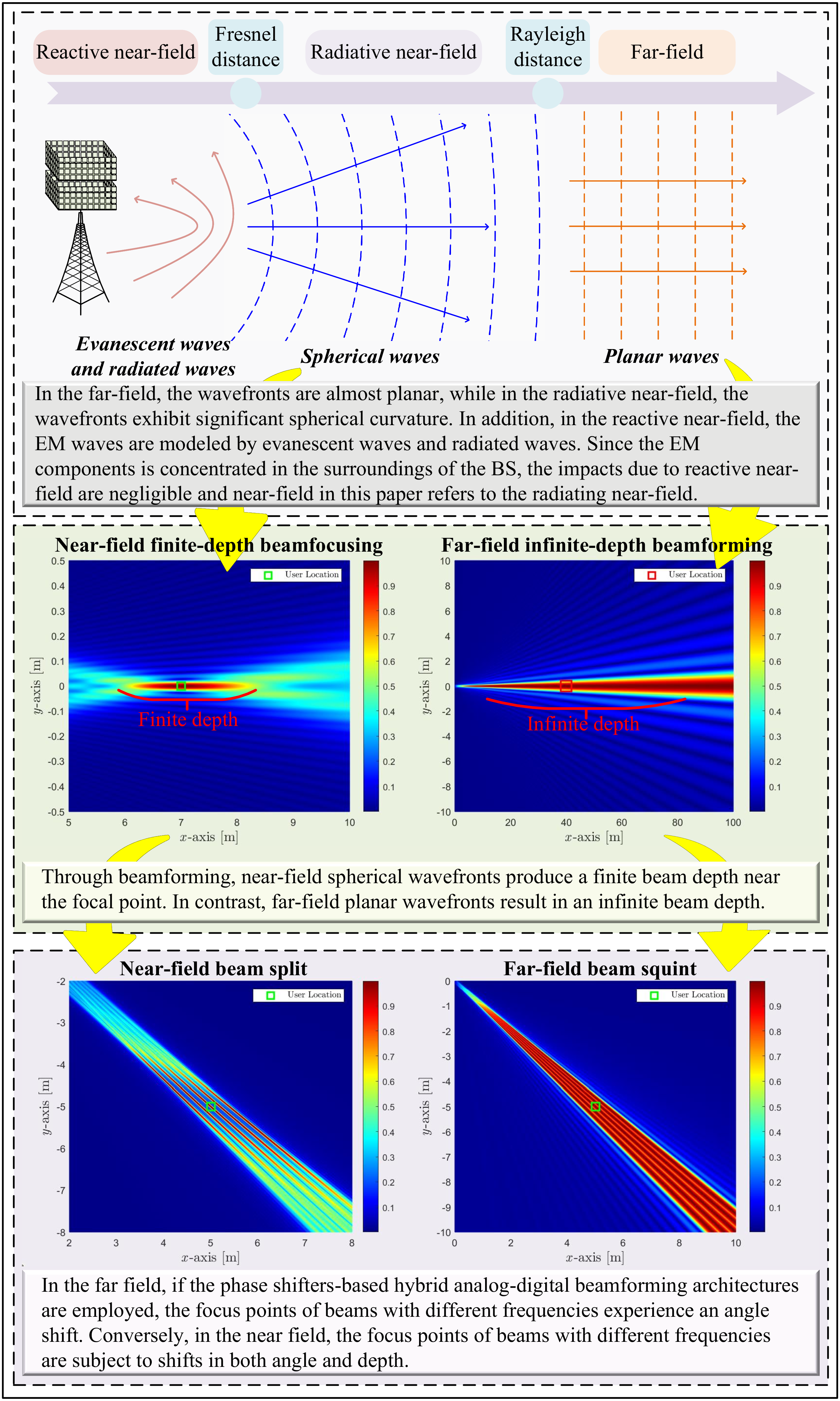}
  \caption{The EM characteristics comparison of near-field and far-field, where the widely adopted boundary is the Fraunhofer/Rayleigh distance. The central carrier frequency is $100$ GHz. We set the number of BS antennas for near-field and far-field scenarios to $512$ and $128$, respectively.}
  \label{EM}
\end{figure}

\subsection{Spherical Wave Properties}

For conventional far-field mMIMO systems, the EM field is often accurately approximated as a plane wave \cite{[1]}-\cite{[2]}.
Under this approximation, the amplitudes and angles of arrival/departure (AoA/AoD) of the transmitted/received signals across all antenna elements are considered identical.
In contrast, near-field XL-MIMO systems necessitate more complex modeling, employing spherical wave assumptions to capture curved phase and amplitude variations across the antenna elements \cite{[1]}-\cite{[2]}.
In such systems, the AoA/AoD and distances between the transmitter and the receiver antennas would vary across the entire array.
Thus, channel modeling must account for both angle and distance information to describe the EM characteristics accurately.

\subsection{Finite-Depth Beamfocusing}

In the far-field, the beam can be steered towards a specific angle through beamforming techniques, as shown in the middle part of Fig.~\ref{EM}.
In contrast, the extra distance information inherent in near-field spherical wavefronts facilitates beamforming to focus the beam at a precise position.
This focusing is characterized by the energy being concentrated at a particular point, with a finite focus area behind it, in both angle and depth.
This is known as finite-depth beamfocusing \cite{[6]} and is shown in the middle figure of Fig.~\ref{EM}. 
While far-field users must be widely spaced in angle to avoid strong interference, finite-depth beamfocusing allows BSs to serve multiple near-field users at the same angle simultaneously but at different distances.  
This capability is a key advantage in near-field communication, enabling more efficient and effective use of the spectrum and spatial resources.

When future networks use phase shifters (PSs) that apply the same phase shift to all subcarriers, it results in the beam squint effects when using wide bandwidth (relative to the carrier frequency).
In traditional far-field mMIMO scenarios, this leads to a small angular shift of the beam, resulting in a minor loss of beamforming gain, as shown in the bottom figure of Fig.~\ref{EM}.
By contrast, in the near-field where the beam is both narrower and has a finite depth, the beam's focus area can be shifted entirely in both angle and depth \cite{[2]}, as shown in the bottom part of Fig.~\ref{EM}.
This is a more severe phenomenon called the \emph{beam split effect}. 
As a remedy, true-time-delay (TTD) units could be implemented to introduce frequency-dependent phase shifts, effectively mitigating beam split effects but at an increased hardware cost \cite{[2]}.

\subsection{Spatial Non-Stationary Properties}

In conventional far-field mMIMO systems, the antenna array aperture of the BS is relatively modest, leading to the average amplitude being spatially stationary among the antennas in the array, even if they vary randomly due to small-scale fading.
Conversely, in near-field XL-MIMO systems, due to near-field scatterers and line-of-sight (LoS) blockages, the physically large array aperture leads to spatial non-stationary channel characteristics \cite{[1]}.
This phenomenon arises because objects of finite size within the propagation environment may no longer function as complete diffuse scatterers for the XL-MIMO array.
The visibility region (VR) concept is widely adopted to characterize these properties \cite{[1]}
and captures which antennas have a LoS path to the scatterer or the UE.
This property might result in the width and depth of the beam becoming wider, leading to a decrease in the actual resolution and spatial multiplexing ability of XL-MIMO systems \cite{[1]}.
An extreme but vivid example is when only a few antennas are visible to the near-field users of the entire array, the working area changes from near-field to far-field, and the beam changes from finite to infinite depth.

\section{Near-Field User Localization and Channel Estimation}

The new EM characteristics introduce several new challenges and opportunities to near-field user localization and channel estimation.
In this section, we first delve into the key challenges that the three new near-field EM characteristics bring to near-field user localization and channel estimation.
Subsequently, we will discuss how to address these challenges based on recent works.
We focus on elucidating the new opportunities posed by these new near-field properties on user localization and channel estimation and exploring strategies to leverage these properties to design more effective user localization and channel estimation algorithms.

\subsection{Near-Field User Localization}

\subsubsection{Challenges}

In the near-field, the unique spherical wave properties allow EM waves to contain both angle and distance information.
This enables signals to be precisely focused on specific locations through beamforming, creating opportunities for a single BS to localize a user fully.
By contrast, traditional localization algorithms designed for far-field scenarios rely on joint efforts by multiple BSs.
For instance, these algorithms typically estimate a user’s location based on the delay or AoA from multiple BSs.
These traditional methods are inefficient in the near-field because they do not effectively utilize the spherical wave characteristics.
Therefore, there is a need to develop new user localization algorithms tailored to spherical wave properties.
Additionally, the spatial non-stationary natures of near-field channels necessitate further innovation in localization algorithm design.
In spatial non-stationary channels, the increased beam depth of near-field beams reduces their beamfocusing precision, and the multiple paths make localization more complex.
Thus, decoupling these channels or designing more efficient algorithms to extract user information effectively under such conditions is crucial.
Moreover, employing hybrid beamforming transceivers in XL-MIMO systems might lead to lower costs and energy consumption, but the beam split effects inherent in these systems must be addressed.
However, if we can manage this frequency-dependent offset, we can potentially search for multiple user locations within a single beam sweeping, offering a new avenue for improving user localization.

\subsubsection{Recent advances}

\begin{table*}[t!]
  \centering
  \fontsize{9}{13}\selectfont
  \caption{Features of the user localization schemes developed for the XL-MIMO systems.}
   \begin{tabular}{
   !{\vrule width1.2pt}  m{1.8 cm}<{\centering}                                                
   !{\vrule width1.2pt}  m{1.9 cm}<{\centering}                         
   !{\vrule width1.2pt}  m{5.9 cm}<{\raggedright}  
   !{\vrule width1.2pt}  m{5.9 cm}<{\raggedright}    
   !{\vrule width1.2pt}  }

    \Xhline{1.2pt}
        \rowcolor{gray!30} \bf Category&  \bf Algorithms &  \bf Algorithm characteristics  &  \bf Remark  \cr\Xhline{1.2pt}

        \multirow{4}{*}{ \makecell{ LoS\\Propagation}}   & Beam training  & {  {Utilizing spherical wave characteristics and finite-depth beam focusing, location estimation can be achieved using array gain as an indicator.}} & { {The use of hierarchical methods, or employing an approach that estimates angle first followed by distance, can effectively reduce costs.}}  \\
        \cline{2-4}  & {  Beam split-assisted beam training}  &  { Exploiting the beam split effect to reduce the cost of beam training.}  &  { Utilizing TTD hardware architectures to generate frequency-dependent phase shifts, effectively controlling the degree of the beam-splitting effect.}  \cr\Xhline{1.2pt}        

        \multirow{5}{*}{ \makecell{Mixed LoS\\and NLoS\\Propagation}}    & { { MUSIC, MLE}}  & Utilizing spherical wave characteristics and employ low-complexity one-dimensional searches to estimate the user's angle and distance continuously. & Decoupling angle and distance to avoid time-consuming binary search. \\
        \cline{2-4}   &{ { DNOMP }}  & Estimating multi-path parameters and deriving user location estimation based on the geometric properties of the paths. & Integrating user localization into the channel estimation process not only reduces additional signal overhead but also enhances the accuracy of both estimations.         \cr\Xhline{1.2pt}

    \end{tabular}
  \vspace{0cm}
  \label{TABLE I}
\end{table*}

Recently, several works have delved into the performance bounds for user localization in XL-MIMO systems \cite{[9]}, \cite{[10]}.
In \cite{[9]}, the authors developed an accurate near-field channel model based on the EM propagation model without approximations.
In \cite{[10]}, leveraging the uniform spherical wave model, the authors derived closed-form expressions for the CRBs for both angle and distance estimation in the near-field XL-MIMO radar mode and phased array radar mode, respectively.
The authors \cite{[10]} proved that the CRB of the angle approaches a limit that depends on the antenna spacing as the array aperture increases and that the CRB of the distance is finite in the near field, unlike in the far field where it is infinite.
These analyses collectively demonstrate the enormous potential of near-field XL-MIMO systems in improving user localization accuracy.

From the perspective of EM propagation and near-field scattering, we next discuss existing user localization algorithms, i.e., LoS propagation and mix LoS and non-line-of-sight (NLoS) propagation. 
In addition, we provide a detailed comparison between recent advances in user localization of XL-MIMO systems in Table \ref{TABLE I}.

$\bullet$ {\emph{LoS Propagation}}

In the near-field, spherical wave properties and finite-depth beamfocusing enable beam training to determine the user's location and angle jointly. 
However, performing a binary search over potential user areas results in significant beam sweeping overhead.
A simple solution to this problem is to use a hierarchical search approach. 
In the first stage, far-field beamforming technology is employed to scan and determine the user's angle. 
In the second stage, based on the obtained angle estimation, near-field beamfocusing technology is used to determine the user's distance.
Another solution builds on the hierarchical search by utilizing the beam split effect. 
TTDs can be used to compensate for beam split effects and control the degree of angle and depth shifts for the beam split \cite{[2]}, \cite{[11]}. 
Specifically, the focus points of the lowest and highest frequency beams can be directed toward the desired angle and distance (i.e., the desired location), indirectly controlling the focus points of beams at other frequencies to cover the desired angle and distance range.
By leveraging the fact that users at different locations receive maximum power at different subcarriers, a low-overhead beam sweeping user localization scheme can be obtained \cite{[11]}.

\begin{figure*}
  \centering
  \includegraphics[width=6.6in]{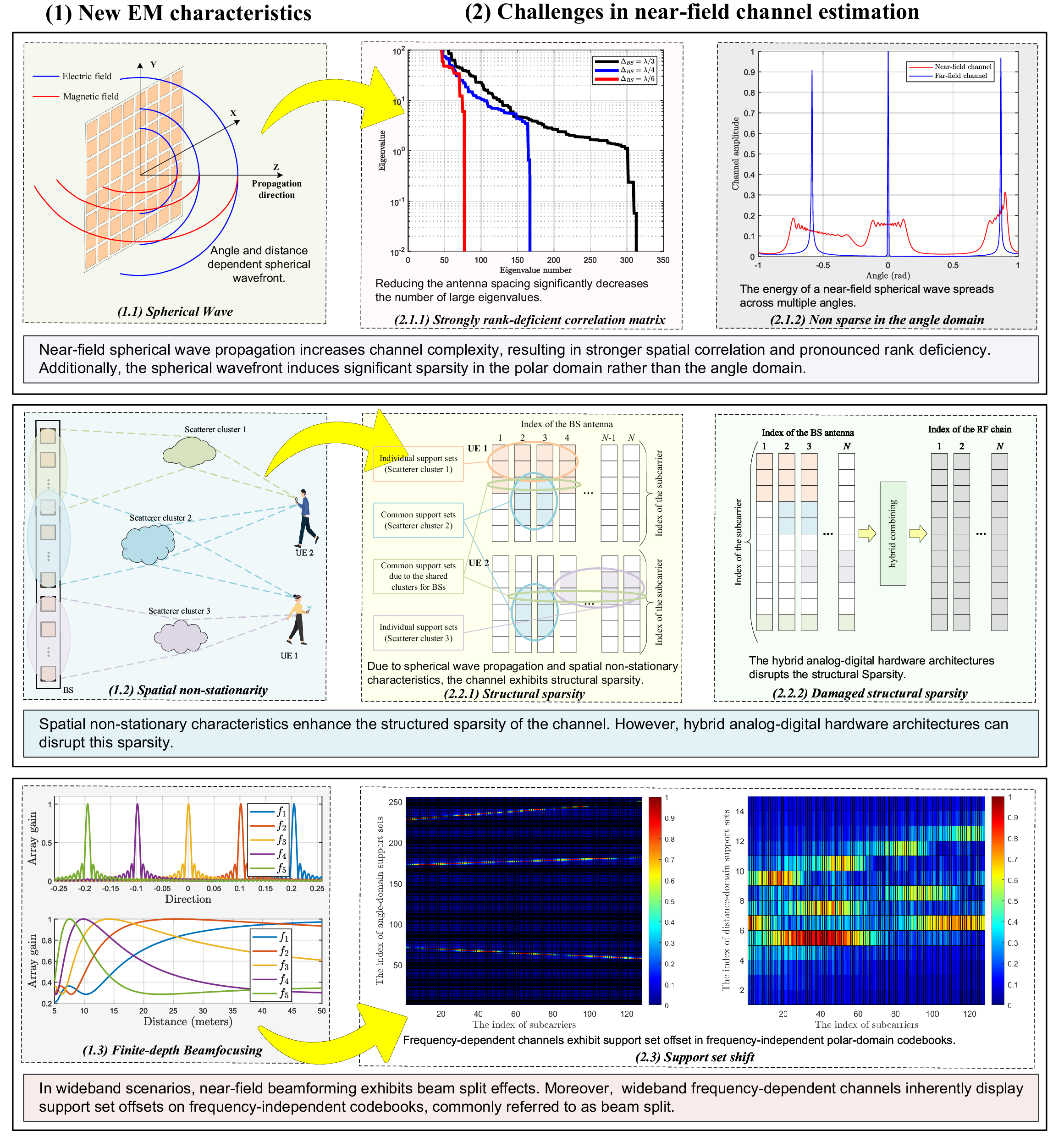}
  \caption{The several new challenges introduced by the three new features of radiative near-field propagation.}
  \label{challenges}
\end{figure*}

$\bullet$ {\emph{Mixed LoS and NLoS Propagation}}

In mixed LoS and NLoS scenarios, several classic algorithms can still be employed to estimate the user's angle and distance from the received signal, such as the multiple signal classification (MUSIC) algorithm and the maximum likelihood estimation (MLE) algorithm. However, when these traditional methods are applied to near-field systems, modifications are necessary to account for the near-field characteristics and to avoid time-consuming two-dimensional (2D) searches.
For instance, in the downlink, the angle can be estimated using the MUSIC algorithm. Subsequently, the distance can be determined through a one-dimensional (1D) distance search with the MLE algorithm \cite{[12]}. The angle and distance estimation can be achieved in the uplink through two consecutive 1D MUSIC algorithms based on the echo signal.
It is important to note that both uplink and downlink localization algorithms continuously estimate the user's angle and distance using low-complexity 1D searches. 
This approach effectively avoids the complex 2D searches required to apply traditional localization algorithms directly.

Notably, an opportunity for near-field communication is to absorb the idea of ISAC, which integrates localization into the channel estimation process. This approach enhances the accuracy of user localization and channel estimation and reduces the additional overhead typically required for user localization in XL-MIMO systems \cite{[13]}.
Specifically, channel estimation algorithms (a damped Newtonized orthogonal matching pursuit (DNOMP) algorithm) can initially estimate the parameters of the LoS and NLoS paths \cite{[13]}. 
The user's location and the complete channel can be determined from these parameters. 
By leveraging the geometric relationship between the user's location and multiple paths, the accuracy and reliability of user localization can be significantly enhanced.

\subsection{Near-Field Channel Estimation}

\subsubsection{Challenges}

Accurate channel estimation is essential for efficiently using all the antennas in XL-MIMO systems.
However, as shown in Fig.~\ref{challenges}, the three new features mentioned above introduce several challenges, which are discussed as follows.
\begin{itemize}
    \item {\bf Strongly rank-deficient channel spatial correlation matrix.} Due to an extremely large number of antennas, the channel spatial correlation matrix is challenging to obtain and exhibits strong rank deficiency characteristics, especially when the antenna spacing is less than half a wavelength \cite{[1]}, \cite{[6]}. This makes traditional minimum mean-squared error (MMSE) schemes impractical for near-field XL-MIMO systems since we cannot afford to estimate the correlation matrices.
    \item {\bf Non-sparse in the angle domain.} 
    In near-field scenarios, each codeword in the codebook must correspond to a spherical wave, associated with both an angle and a distance \cite{[1]}, \cite{[2]}. Hence, near-field channels are no longer sparse in the angle domain, necessitating much larger codebooks for effective channel estimation.
    \item {\bf Structural sparsity.} The spatial non-stationary characteristics also introduce new sparsity patterns \cite{[15]}, \cite{[16]}. Specifically, instead of all antennas in the BS serving specific users, only the antennas within the VR of each user are active, resulting in sparsity in the antenna domain \cite{[16]}. Moreover, channels within the same VR exhibit common sparsity patterns. This new form of sparsity provides a foundation for designing advanced channel estimation algorithms. 
    \item {\bf Damaged structural sparsity.} In addition, considering the hybrid beamforming transceivers, when the number of radio-frequency (RF) chains in the BS is less than the dimension of the received signals, the received signals are a mixture of signals from all antennas. This disrupts the sparsity in the antenna domain \cite{[15]}, posing additional challenges for channel estimation schemes. The issue becomes larger in wideband scenarios due to the beam split effect  \cite{[18]}. 
\end{itemize} 

\subsubsection{Recent advances}

\begin{table*}[t!]
  \centering
  \fontsize{9}{13}\selectfont
  \caption{Features of the channel estimation schemes developed the XL-MIMO systems.}
   \begin{tabular}{
   !{\vrule width1.2pt}  m{1.8 cm}<{\centering}                                                
   !{\vrule width1.2pt}  m{2.1 cm}<{\centering}                         
   !{\vrule width1.2pt}  m{5.7 cm}<{\raggedright}  
   !{\vrule width1.2pt}  m{6.1 cm}<{\raggedright}     
   !{\vrule width1.2pt}  }

    \Xhline{1.2pt}
        \rowcolor{gray!30} \bf Category&    \bf Algorithms &  \bf Algorithm characteristics  &  \bf Remark  \cr\Xhline{1.2pt}

        \multirow{9}{*}{ \makecell{ Spherical \\wave-driven\\ schemes}}   & RS-LS  & {  { Exploit the spatial correlations resulting from the array geometry and general propagation environment characteristics.}} & { {Utilize a reduced-rank subspace of the spatial correlation matrix to reduce complexity and achieve performance close to MMSE.}}  \\
        \cline{2-4}   & { { P-SOMP, P-SIGW}}  &  { Exploit the polar-domain sparsity of near-field channels.}  &  { Prove the sufficient sparsity of near-field channels in the polar domain.}  \\
        \cline{2-4}   &{ { on-grid SGP, off-grid SGP}}  & { Further develop the sparsity of near-field channels with lower complexity.}  & { Exploit an alternating minimization approach to refine channel estimates.}   \\
        \cline{2-4}   &{ { DPSS-based scheme}}  & Utilize the DPSS-based codebook to replace traditional DFT-based and spherical wave-based codebooks for the CS-based channel estimation.  & Reduce dictionary size by 88\% by utilizing the inherent orthogonality between the channel’s feature vectors, compared to the DFT and spherical wave-based dictionaries.  \cr\Xhline{1.2pt}        

        \multirow{5}{*}{ \makecell{Spatial non-\\stationary-\\driven\\ schemes}}  & { { Turbo-OAMP}}  & Utilize the LBP instead of the non-linear MMSE module in the traditional OAMP algorithm to more effectively exploit structured sparsity. & Utilize three layers of Markov chains to model the structured sparsity for spatial non-stationary properties. \\
        \cline{2-4}  &{ { G-P-SOMP, G-P-SIGW}}  & Firstly, transform a spatial non-stationary channel into several spatial stationary subchannels, and then perform channel estimation on each subchannel. & Exploit the GTBC to construct correlations for subchannels to achieve the ability to identify spatial non-stationary information.         \cr\Xhline{1.2pt}

        {\multirow{4}{*}{\makecell{Finite-depth\\beamfocusing-\\driven\\schemes}}}  &  { { BPD-based scheme}}   & First calculate the support set offset at different frequencies and then exploit frequency-independent polar-domain codebooks to recover the channel. &   \\
        \cline{2-3}   & { { NBA-OMP, NBA-OMP-FL}}  & Directly employ frequency-dependent polar-domain codebooks for channel recovery. &   \multirow{-5}{*}{\begin{tabular}[c]{@{}l@{}}Exploit the prior known offsets of the support \\ sets in the angle and distance domains at \\ different frequencies to compensate for \\ the impact of beam split  effects.\end{tabular}}  \cr\Xhline{1.2pt}

    \end{tabular}
  \vspace{0cm}
  \label{TABLE II}
\end{table*}

Some recent works have been devoted to addressing the challenges.
In this part, we will introduce three categories of XL-MIMO channel estimation schemes from the perspective of utilizing the characteristics of XL-MIMO systems.

$\bullet$ {\emph{Spherical Wave-Driven Schemes}}

Traditional model-agnostic estimation schemes, such as the least-squares (LS) method, remain applicable to near-field communication but do not exploit the special near-field characteristics.
On the other hand, the MMSE scheme is seldom adopted in XL-MIMO systems due to the complexity involved in obtaining the spatial correlation matrix for each mobile UE.
Building on the concept of utilizing the spatial correlation matrix to minimize MSE as in MMSE, the authors in \cite{[6]} proposed a reduced-subspace LS (RS-LS) scheme.
This method leverages user-independent spatial correlation properties resulting from the array geometry and general propagation environment characteristics.
The RS-LS scheme outperforms the LS scheme, which also does not require user-specific statistical information while achieving performance close to that of MMSE but with significantly lower complexity.

In situations where the number of transmit antennas is large but the channel is known to contain a small number of dominant paths, the pilot overhead can be reduced using compression sensing (CS)-based algorithms.
The core of classical algorithms lies in leveraging the sparsity in the angle domain under the assumption of plane waves. 
However, if traditional Fourier codebooks are used to transform near-field channels into the angle domain, the energy of each near-field path will be spread across multiple angular bins \cite{[2]}.
To address this issue, the authors in \cite{[2]} proposed a larger polar-domain codebook that simultaneously considers distances and angles, allowing both near-field and far-field paths to exhibit sparsity in the polar domain. 
Leveraging this polar-domain sparsity, a polar-domain simultaneous orthogonal matching pursuit (P-SOMP) scheme and a polar-domain simultaneous iterative gridless weighted (P-SIGW) scheme was introduced to estimate channels containing a small number of spherical wave paths \cite{[2]}. 
Similarly, the authors in \cite{[21]} explored the sparsity of near-field channels in the polar domain and its application in channel estimation. 
Specifically, both on-grid and off-grid stochastic gradient pursuit (SGP)-based schemes were proposed for hybrid far- and near-field (cross-field) communications. 
These algorithms demonstrate the effectiveness of polar-domain CS algorithms in near-field channel estimation.

Interestingly, user localization can significantly enhance channel estimation performance in near-field XL-MIMO systems. 
By utilizing eigenvalue decomposition of near-field channels, a novel discrete prolate spheroidal sequences (DPSS)-based codebook was derived \cite{[14]}. 
This codebook's primary advantage is its compact size, resulting from the inherent orthogonality between the feature vectors of the channel. 
However, a notable disadvantage is that the codebook depends on the user's location coordinates.
Building on this, a sensing-enhanced channel estimation method has emerged \cite{[14]}. 
This method first estimates the user's location using a time reversal algorithm. The location estimation is then used to generate the DPSS-based codebook. Finally, the channel is recovered using a CS-based algorithm. 
Notably, due to the orthogonality of the channel's feature vectors, the size of the DPSS-based codebook is smaller than that of the polar-domain codebook while achieving higher channel estimation accuracy.

$\bullet$ {\emph{Spatial Non-Stationary-Driven Schemes}}

Estimation algorithms developed for spatially stationary channels can perform badly if the channel features significant spatial non-stationary. 
Effectively capturing and utilizing the structural sparsity caused by these non-stationary characteristics is crucial for designing efficient near-field channel estimation algorithms.
One promising method involves using a Bayesian factor graph based on hidden Markov model (HMM) priors to capture the structural sparsity in the antenna-delay domain under spatial non-stationary conditions \cite{[16]}. 
Specifically, the VR and delay domain clustering are modeled using three layers of Markov chains. 
Subsequently, a turbo orthogonal approximate message passing (Turbo-OAMP) framework is introduced to estimate the non-stationary channel parameters \cite{[16]}. 
Compared to traditional OAMP, this approach incorporates loopy belief propagation to achieve approximate Bayesian inference, enabling a deeper utilization of spatial non-stationary characteristics.

Another challenge posed by spatial non-stationary characteristics arises when using a hybrid beamforming transceiver. 
Different paths from a user will traverse different VRs, and all these signals converge at the BS, causing VR coupling. 
Additionally, the received signal at each RF chain is a composite of the signals from all antennas, further compounding the coupling of sparse characteristics.
To address this, a group time block code (GTBC)-based method can be employed to create correlations in the time domain, enabling the identification of spatial non-stationary characteristics within the coupled signal \cite{[15]}. 
This approach transforms a spatially non-stationary channel into several spatially stationary subchannels. 
Subsequently, CS-based algorithms, such as the P-SOMP and P-SIGW, can be utilized to estimate these spatially stationary subchannels.

$\bullet$ {\emph{Finite-Depth Beamfocusing-Driven Schemes}}

There are two solutions to address the challenges posed by finite-depth beamfocusing, particularly the support set shift caused by beam split effects.
One solution is to estimate subcarriers of different frequencies independently. 
This approach involves performing channel estimation based on frequency-specific codebooks tailored to each frequency with CS algorithms. 
While effective, this method is highly complex.
An alternative solution is calculating the degree of support set offset of a specific frequency relative to the center carrier frequency based on the beam split effect. 
According to the mutual coherence of near-field array response vectors at different frequencies, the sparse support sets in both the angle and distance domains grow linearly with the subcarrier frequency, a phenomenon known as the bilinear pattern \cite{[22]}.
Building on this, a support set offset matrix can be constructed with all angle and distance samples at different frequencies. 
This matrix corrects the support set offset caused by beam split effects when using polar-domain codebooks to recover channels at different frequencies. 
The bilinear pattern detection (BPD)-based CS algorithm effectively compensates for the impact of beam split on channel estimation, significantly improving estimation accuracy \cite{[22]}.
The authors in \cite{[18]} applied this same approach to propose a near-field beam-split-aware orthogonal matching pursuit (NBA-OMP) scheme, further combining it with federated learning (FL) to reduce computational complexity.

To sum up, a detailed comparison of the recent channel estimation schemes is presented in TABLE \ref{TABLE II}.

\section{Case Studies}
\label{Case}

In this section, we introduce near-field user localization and channel estimation designs for XL-MIMO systems. 
We propose specific solutions that have not been addressed in previous works.

\subsection{Near-Field User Localization}

\begin{figure}
  \centering
  \includegraphics[width=3.3in]{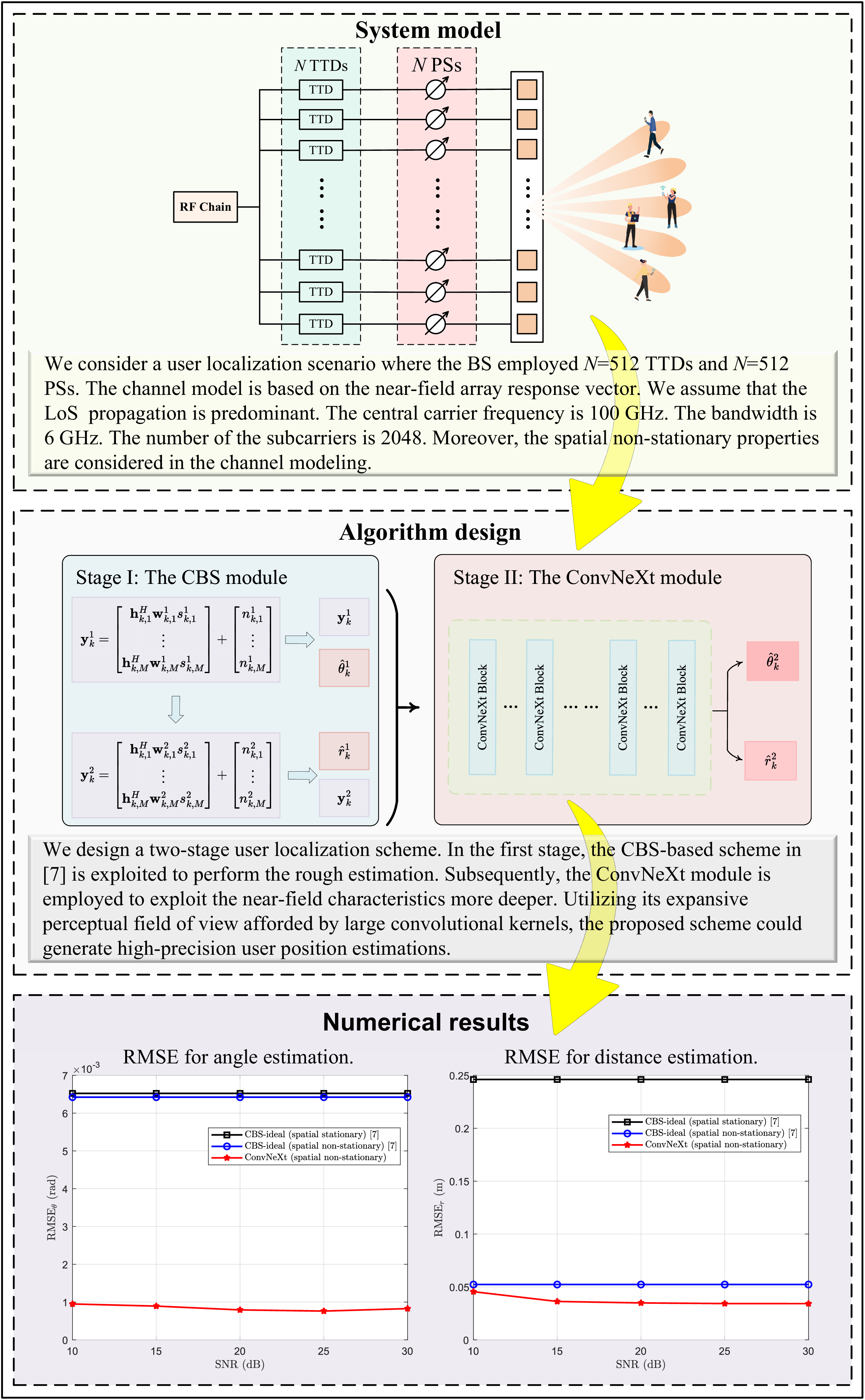}
  \caption{User localization in wideband XL-MIMO systems with analog beamforming transceivers, where the BS is equipped with a uniform linear array (ULA) to serve four users.}
  \label{localization}
\end{figure}

In user localization, we consider a scenario incorporating spherical wave characteristics, spatial non-stationary properties, and beam split effects, as depicted in Fig.~\ref{localization}. 
For the root mean square error (RMSE) performance comparison of various localization schemes, we include the controllable beam split (CBS)-based scheme from \cite{[11]}. 
Specifically, the CBS is evaluated in an ideal, noise-free environment (denoted as CBS-ideal), assuming that the users' angles are known during the distance estimation stage.
The proposed deep learning (DL)-based scheme aims to employ a convolutional neural network model (ConvNeXt), utilizing its expansive perceptual field of view afforded by large convolutional kernels, to learn noise characteristics and the connections and mappings between the rough estimation information and different subcarriers. 
The first observation is that the non-stationary characteristics impact distance estimation accuracy more than angle estimation accuracy. 
This is because the non-stationary nature of space increases the depth of the near-field beam, leading to lower resolution for distance estimation. 
Notably, while angle estimation can utilize far-field beams, distance estimation relies solely on near-field beams.
Additionally, by performing channel estimation under ideal conditions, we have demonstrated the potential for a single BS to estimate user locations in near-field systems independently. 
Furthermore, we achieved centimeter-level user localization by employing DL techniques even in noisy environments.

\subsection{Near-Field Channel Estimation}

\begin{figure}
  \centering
  \includegraphics[width=3in]{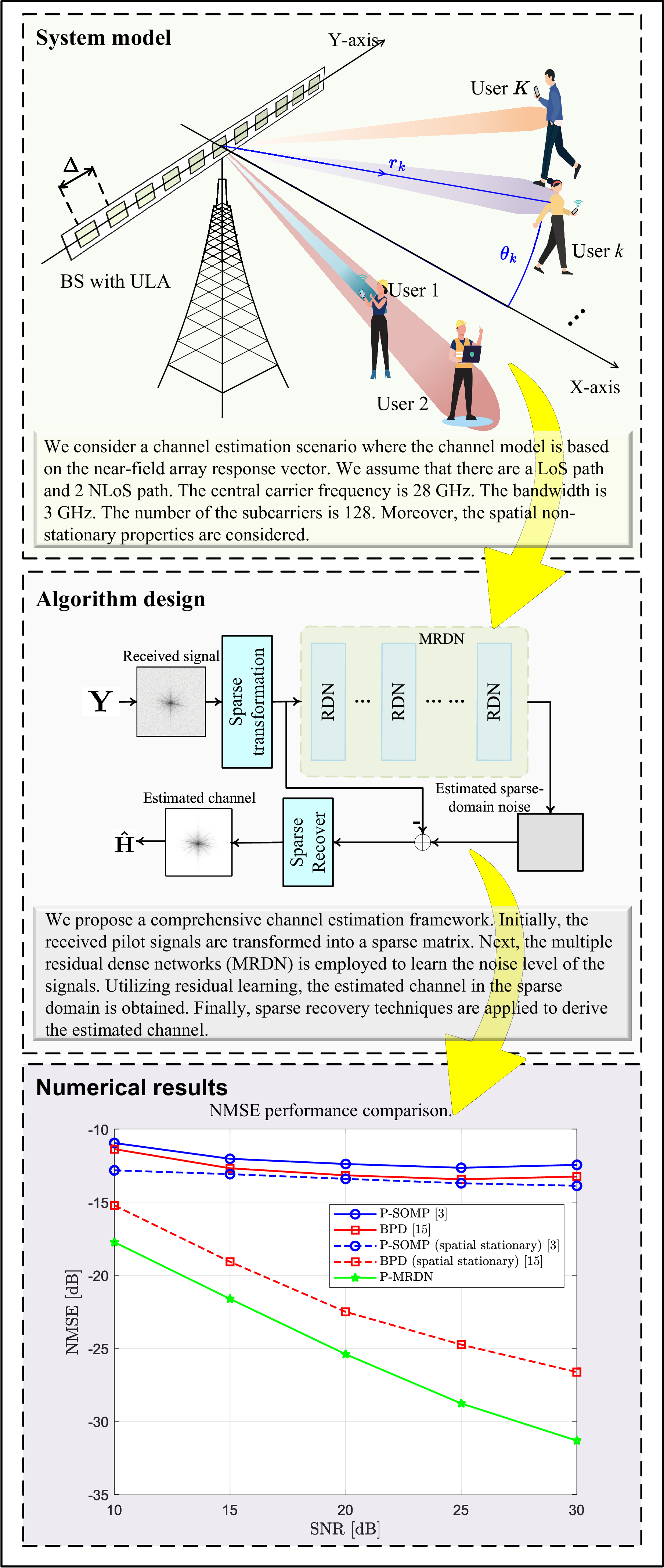}
  \caption{Channel estimation in wideband XL-MIMO systems, where the BS is equipped with a uniform linear array (ULA) to serve multiple users.}
  \label{estimation}
\end{figure}

As shown in Fig.~\ref{estimation}, we consider a channel estimation scenario where multiple UEs send orthogonal uplink pilots. 
The first notable finding is that in wideband XL-MIMO scenarios, the BPD-based scheme in \cite{[22]}, which accounts for support set offset correction, outperforms the P-SOMP scheme in \cite{[2]}. Additionally, we observe that spatial non-stationary characteristics significantly impact the normalized MSE (NMSE) of both BPD and P-SOMP. 
This is because these schemes rely on polar-domain codebooks without considering spatial non-stationary effects.
More importantly, the polar-domain MRDN-based scheme achieves the highest NMSE performance. 
By leveraging techniques such as DL, residual learning, sparse transformation, and multi-scale feature extraction, the MRDN-based scheme effectively mitigates the effects of beam split and spatial non-stationary characteristics, resulting in the most accurate channel estimation.

\section{Future Directions}
\label{Future}

\subsection{Refined Electromagnetic Models}

The near-field channel models commonly used for analytical and algorithmic development are simplifications assuming far-field propagation between any pair of transmit and receive antennas, but noticeable non-planar phase variation across different antennas in an array. More precise models can be developed to capture e.g., amplitude variations and mutual coupling within an array. Coupling between the transmitter and receiver can also occur in some situations because the reactive near-field also grows with the array dimensions and carrier frequency. It remains to be determined how detailed models are needed for algorithmic design and for accurate performance analysis, respectively. An industry-standard for near-field channel simulation must be developed, based on measurements, ray-tracing, and analytical formulas. Machine learning methods can possibly be used to find the most prominent features to model in real near-field channels. 

\subsection{Hardware Distortion}

The wireless channel between a transmitter and receiver is not only determined by wave propagation but the transceiver hardware and array geometry are intrinsic parts. The aforementioned algorithms exploit the near-field-specific channel characteristics under the assumption that the hardware is ideal and array geometry is known. None of these conditions can be ensured in practice. Firstly, the array response vector function can differ from its theoretical expression due to mismatches in the antenna locations, non-uniform antenna properties, etc. Moreover, non-linear amplification, imperfect filters, and phase noise can degrade the algorithmic performance. We envision that future algorithmic development, possibly with trainable parameters, can mitigate these issues and exploit its characteristics to enable new features---as was the case when the beam split effect was addressed.

\subsection{Electromagnetic Information Theory}

The main motivation for XL-MIMO is to manage more data traffic than in current networks. Hence, the performance metric is the capacity obtained by combining electromagnetic channel modeling with information-theoretic methodologies. The maximum spatial DoFs that a planar XL-MIMO array can handle have been characterized as $\pi \cdot \textrm{Area} / \textrm{wavelength}^2$, but this is only a first-order capacity approximation describing how many non-zero singular values a channel matrix can have. Future research must determine what array configurations give the highest average capacity for a given user population, thereby shaping individual singular values. It is particularly important to study the interplay between array shape, propagation environment, and capacity in the upper mid-band, which will play a central role in 6G.

\section{Conclusions}
\label{Conclusions}

When XL-MIMO technology is used in future networks, the algorithms for localization and channel estimation can be refined to capture the three distinctive near-field properties: spherical waves, beamfocusing, and spatial non-stationarity. This article has summarized the related challenges and recent advances, and highlighted three esential research directions for future work.

\bibliographystyle{IEEEtran}
\bibliography{IEEEabrv,Ref}
\end{document}